\documentstyle[preprint,aps,prl]{revtex}
\begin{document}
\draft
\title{Color singlet suppression of quark-gluon plasma formation}
\author{Jes Madsen, Dan M. Jensen and Michael B. Christiansen}
\address{Institute of Physics and Astronomy,
University of Aarhus, 
DK-8000 \AA rhus C, Denmark}
\date{June 7, 1995}
\maketitle

\begin{abstract}
The rate of quark-gluon plasma droplet nucleation in superheated hadronic
matter is calculated within the MIT bag model. The requirements of color
singletness and (to less extent) fixed momentum
suppress the nucleation rate by many orders of magnitude, making
thermal nucleation of quark-gluon plasma droplets unlikely in
ultrarelativistic heavy-ion collisions if the transition is first order
and reasonably described by the bag model.
\end{abstract}

\pacs{25.75.+r, 12.38.Mh, 12.39.Ba, 24.85.+p\\ \\ \\
To appear in PHYSICAL REVIEW C, April 1996} 

Ultrarelativistic heavy-ion collision experiments at CERN
and Brookhaven aim at the formation of quark-gluon plasma.
No unambiguous signals have been detected so far, but the prospects will
improve tremendously with the next generation of colliders (LHC and
RHIC). Normally it is
assumed that formation of a quark-gluon plasma will take place if a
region in the colliding nuclei reach a combination of temperature and
chemical potential that brings it into that part of the phase diagram
for hadronic matter where the quark-gluon plasma has a lower free energy
than a hadron gas. 

Should the phase transition be of first order,
the rate for quark-gluon droplet formation, $\cal R$, can be estimated in the
framework of homogeneous nucleation theory\cite{landau}, where
\begin{equation}
{\cal R}\approx T^4 \exp(-\Delta F/T)\approx 0.2
{\rm fm}^{-3}(10^{-23}{\rm s})^{-1}T_{100}^4\exp(-\Delta F/T).
\label{rate}
\end{equation}
Here $\Delta F$ is the height of the free energy barrier which the thermal
nucleation has to overcome, $T_{100}$ is the temperature in units of
100MeV, and the rate has been expressed in units of typical heavy ion
collision volumes and time scales. There is an extensive literature on
the proper choice of prefactor, but since the exponential is by far the
most important for our present discussion (the prefactor does not
influence the relative rate suppression due to the effects discussed in
our paper), we have chosen the
dimensional estimate $T^4$. We note that a more realistic choice of
prefactor\cite{venugopalan} could change the nucleation rate by a few orders
of magnitude\cite{kapusta}. A quantitative estimate of the effect in the
present context is hindered by the fact that no derivation of the
prefactor exists for a situation where curvature rather than surface
tension is the most important contribution to the surface energy.

It is normally assumed that a moderate superheating is sufficient to
allow droplet nucleation. This is confirmed by calculations of the
droplet free energy in simple phenomenological models like, e.g., the
MIT bag model\cite{degrand}. Such calculations\cite{kapusta} have, 
however, neglected the
requirements of color singletness and fixed momentum for the quark
droplet. Both of these constraints significantly reduce the effective degrees of
freedom in the quark-gluon plasma\cite{elze}, thereby increasing the free energy
barrier, $\Delta F$, and reducing the nucleation rate. As we show below,
the nucleation rate is in fact reduced by many orders of magnitude,
making thermal nucleation virtually impossible regardless of the
amount of superheating.

We note that the ingredients of our rather simple calculation have
been around in the literature for about a decade, but to our knowledge
no-one has checked these dramatic consequences explicitly before.

We want to calculate the nucleation rate self-consistently within the
MIT bag model, including two flavors of massless quarks,
gluons, a bag pressure, volume and curvature
energies, and the constraint that
quark droplets must have a fixed momentum and be color neutral.
It will turn out that our conclusions are not very sensitive to chemical
potential, bag constant, number of quark flavors or the inclusion of
the hadron gas (in most of the following we therefore just ascribe a
chemical potential but no pressure to the hadrons), 
but very sensitive to the color singlet constraint.

Elze and Greiner\cite{elze} have derived the color singlet
fixed-momentum grand canonical partition function, $Z(T,R,\mu;p)$,
for a hot, spherical quark-gluon plasma droplet of temperature $T$, radius $R$,
quark chemical potential $\mu$ (corresponding to a baryon chemical
potential $\mu_B=3\mu$), and total momentum $p$. From the partition
function one can find the grand potential, $\Omega =-T\ln Z$, as
\begin{equation}
\Omega =T\left[ \ln(2\pi\sqrt{3})+4\ln C\right] +1.5T\left[\ln D -\ln
\pi\right] +\frac{p^2}{T4D} +BV-T\ln Z_0 ,
\label{free}
\end{equation}
where
\begin{equation}
\ln Z_0\equiv X-Y,
\end{equation}
\begin{equation}
D\equiv 2X-\frac{1}{3}Y,
\end{equation}
\begin{equation}
X\equiv
\frac{\pi^2}{12}VT^3\left\{12N_Q\left[\frac{7}{60}+ 
\frac{1}{2}\left(\frac{\mu}{\pi T}\right)^2+
\frac{1}{4}\left(\frac{\mu}{\pi T}\right)^4\right]+\frac{32}{15}\right\} ,
\end{equation}
\begin{equation}
Y\equiv
\frac{\pi}{18}RT\left\{12N_Q\left[\frac{1}{2}+ 
\frac{3}{2}\left(\frac{\mu}{\pi T}\right)^2\right]+64\right\} ,
\end{equation}
\begin{equation}
C\equiv
\frac{1}{6}VT^3\left\{2N_Q\left[1+
3\left(\frac{\mu}{\pi T}\right)^2\right] +12\right\}+
\frac{1}{3\pi}RT\left( 24-2N_Q\right) .
\end{equation}
$N_Q$ is the number of massless quark flavors (we take $N_Q=2$ in most
of the following). $V$ is the volume of the droplet. The first term in
the grand potential is the contribution from the color singlet projection.
The next two terms are from fixing the momentum (notice that the first
of these terms survive even for $p=0$), and the last two terms are,
respectively, the bag energy proportional to the bag constant, $B$, and
the ``normal'' bag model grand potential of the quark-gluon gas. Notice that
finite size effects, i.e.\ energy terms proportional to the extrinsic
curvature ($8\pi R$) of the droplet have been included, whereas massless quarks
and gluons contribute no surface tension term proportional to the bag
area. One could add to Eq.\ (\ref{free}) a surface energy from the
hadron phase, as well as the hadron pressure times volume
($\Omega\mapsto\Omega+P_{\rm Hadron}V+\Omega_{\rm Surface,Hadron}$,
where $P_{\rm Hadron}$ is the pressure of the hadron phase).
These terms would both add positive contributions to
$\Omega$, thereby further decreasing the quark-gluon droplet formation
rate. For reasonable choices of parameters these terms are not decisive,
and since our aim is to clearly demonstrate the enormous rate
suppression due to the color singlet and momentum constraints, we have
chosen not to incorporate them. Had we done that, they would only
further suppress the nucleation rate and strengthen our conclusion (we
return to this issue below).

The partition function was derived in a saddle-point approximation, which is
expected to break down for $RT\rightarrow 0$. When $\mu=0$ 
it should be good to 30\%
for $RT=1$ and a few percent for $RT\approx 2$, and the
error decreases rapidly for $\mu >0$\cite{elze}.

The grand potential as a function of droplet radius
for fixed bag constant, $\mu=0$, and a temperature corresponding to 
nearly 30MeV superheating
is shown in Figure 1. Curves are given with and without
inclusion of the color singlet and $p=0$ constraints.
One notes a significant increase in $\Omega$, in particular due to the
color singlet requirement.

In Figure 2 we show the corresponding nucleation rate of quark-gluon
droplets estimated according to Eq.~(\ref{rate}). The change in free energy,
$\Delta F$, necessary for formation of a critical bubble in chemical
equilibrium with the surrounding hadrons at fixed temperature
is just the height of the $\Omega$ barrier, so $\Delta F=\Omega$ with
$\Omega$ given in Eq.\ (\ref{free}). 
One recognizes the well-known reduction of the
transition temperature for increased chemical potential, but what is
more important in the present context is the significant suppression (by
4--5 orders of magnitude at extreme superheating, and much more at
moderate superheating) of the nucleation rate. Units in the plot are
chosen such that unity corresponds to 1 nucleation per fm$^3$ per
$10^{-23}$s (=3 fm/c). Typical volumes and time scales relevant for
quark-gluon plasma formation in ultrarelativistic heavy-ion collisions
could amount to maybe 100 fm$^3\cdot$fm/c, 
so in standard calculations\cite{kapusta}
one predicts a fair probability for bubble nucleation in the case of,
say, a 10MeV superheating. However, the suppression resulting from
requiring fixed momentum and color singletness gives a very low
probability, even for (unrealistic?) superheatings of 50--100 MeV.

This surprising conclusion is not significantly influenced by changes in
the assumed number of massless quarks (here taken to be 2, corresponding
to infinite s-quark mass), or to the external pressure contributed by
the hadron gas (the latter corresponds effectively to a small increase
in $B$\cite{hadrons}). Other choices of $B$ would correspond
to a rescaling of $T$ and $\mu$ in Fig.\ 2 proportional to $B^{1/4}$,
and of $\cal R$ proportional to $B$ (i.e., no qualitative change in the
conclusion). A massive s-quark would add a surface tension term to the
energy, and a surface tension contribution could come from the hadron
phase as well. This would change the numbers (always in the direction of
even lower nucleation probability and a slightly higher critical radius), 
but not the effect of color singlet
suppression. No derivation exists of the color singlet partition
function for a massive quark; neither does a calculation for non-zero
strong coupling constant, which has therefore been assumed equal to zero
in the present investigation. 

The results do of course depend on the choice of model for the
quark-gluon plasma. The MIT bag is certainly relevant only if a first
order phase transition is involved at all. Some lattice calculations
indicate, that the transition could be second order for $\mu=0$, but
the issue is far from settled, and calculations for $\mu\neq 0$ do not
exist. Other models would give other numerical
suppression factors, but the color singlet constraint will in any case
reduce the effective number of degrees of freedom \cite{elze}, thereby
increasing $\Delta F$ in any model, so qualitatively a rate suppression
should be expected.

Equation (\ref{free}) is based on a saddle-point approximation, which is
again based on an expansion of the density of states in terms of volume
and curvature terms. The expansion is known to reproduce direct
summation very well for $\mu=0$\cite{mardor} and for $T=0$
\cite{madsen}, so there is no reason to disbelieve this assumption.
The uncertainty in the saddle-point approximation was discussed above.
It is not negligible (but not devastating either) for critical bubbles
with $RT\approx 0.9$ as typically found for superheating beyond 20MeV
when $\mu=0$,
but negligible for the much larger bubbles involved closer to the bulk
phase transition temperature, and for $\mu\gg 0$ where the saddle-point
approximation is much better\cite{elze}. Figure 3 shows the radius of critical
bubbles as a function of temperature.

Thus, in spite of all the reservations, we conclude that a very
significant suppression of the nucleation rate for quark-gluon plasma
droplets in a superheated hadron gas is an inevitable consequence of the
fixed momentum and color singlet constraints if the quark-hadron phase
transition is first order. Other mechanisms than thermal nucleation
(e.g., nucleation due to impurities) are needed to form quark-gluon
plasma droplets in ultrarelativistic heavy-ion collisions if the phase
transition is first order.

\begin{figure}
\caption{The grand potential, $\Omega$,
as a function of droplet radius for $B^{1/4}=200$MeV, $\mu=0$, and $T=170$MeV.
The lower curve is without momentum and color singlet
constraints, middle curve with the color singlet constraint, and upper
curve with both constraints included (at zero momentum).}
\label{fig1}
\end{figure}
\begin{figure}
\caption{The nucleation rate of quark-gluon droplets as a function of
temperature for $B^{1/4}=200$MeV. Solid curves for quark chemical
potential zero, dotted curves for $\mu=100$MeV, dashed curves for
$\mu=300$MeV, dot-dash curves for $\mu=400$MeV, and dash-triple dot
curves for $\mu=500$MeV. The lower set of curves includes the color and momentum
constraints; the upper set does not.}
\end{figure}
\begin{figure}
\caption{Radius of critical bubbles as a function of temperature for
calculations including color and momentum constraints, for chemical
potentials as in Fig.\ 2.}
\end{figure}


\begin{references}
\bibitem{landau} L.\ D.\ Landau and E.\ M.\ Lifshitz, {\it Statistical
Physics} (Pergamon, New York, 1980).
\bibitem{venugopalan} R.\ Venugopalan and A.\ Vischer, Phys.\ Rev.\ E
{\bf 49}, 5849 (1994); L.\ P.\ Csernai and J.\ I.\ Kapusta, Phys.\
Rev.\ D {\bf 46}, 1379 (1992).
\bibitem{kapusta} See for example 
J.\ I.\ Kapusta, A.\ P.\ Vischer, and R.\ Venugopalan,
Phys.\ Rev.\ C {\bf 51}, 901 (1995).
\bibitem{degrand} T.\ A.\ DeGrand, R.\ L.\ Jaffe, K.\ Johnson, and J.\
Kiskis, Phys.\ Rev.\ D {\bf 12}, 2060 (1975).
\bibitem{elze} H-Th. Elze and W. Greiner, Phys.\ Lett.\ B {\bf 179}, 385
(1986).
See also H-Th. Elze, W. Greiner, and J.\ Rafelski, Phys.\ Lett.\ {\bf 124B},
515 (1983); Z.\ Phys.\ C {\bf 24}, 361 (1984). We note that other
expressions for the color singlet partition function (without a momentum
constraint) deviating from that of Elze and Greiner have been given by 
M.\ I.\ Gorenstein, S.\ I.\ Lipskikh, V.\ K.\ Petrov, and G.\ M.\
Zinovjev, Phys.\ Lett.\ {\bf 123B}, 437 (1983), and M.\ G.\ Mustafa,
Phys.\ Lett.\ B {\bf 318}, 517 (1993). We have successfully reproduced
the calculation of Elze and Greiner.
\bibitem{hadrons} The effect of the hadronic pressure can be calculated
by substituting $B+P_{\rm Hadron}$ instead of $B$ in Eq.\ (\ref{free}).
For $\mu=0$ an (over)estimate of $P_{\rm Hadron}$ is the pressure of a
massless pion gas with 3 degrees of freedom, $P_\pi =\pi^2T^4/30$. The
results with a hadron gas thus corresponds to those without a hadron
gas for bag constant $B^{1/4}\left[1+0.33(T/B^{1/4})^4
\right]^{1/4}$ instead of $B^{1/4}$. For temperatures near the phase
transition ($T\approx 0.7B^{1/4}$) this is less than a 2\% effect on
the scaling of $B^{1/4}$ and correspondingly on radii, temperatures etc.
In the other extreme of very high baryon density and low $T$ the leading
contribution to the hadron pressure can be approximated by a
parametrisation like the Bethe-Johnson equation of state including a
repulsive core via $\omega$-exchange (H.\ A.\ Bethe and M.\ B.\ Johnson,
Nucl.\ Phys.\ A {\bf 230}, 1 (1974)). The simplest case for a pure
neutron gas is typical for more sophisticated calculations, and gives
$P_{\rm Hadron}=364n^{2.54}$MeV fm$^{-3}$, where $n$ is the baryon
density per fm$^3$. As for $\mu=0$ one can use the calculations without the
hadron pressure by substituting instead of $B^{1/4}$ the expression
$B^{1/4}\left[ 1+\beta\right]^{1/4}$, where 
$\beta=0.017((200{\rm MeV})^4/B)(n/n_{\rm nuc})^{2.54}$,
and $n_{\rm nuc}$ is nuclear matter density. The corresponding changes
in radii etc are small unless one goes to very
extreme densities (0.4\% for $n=n_{\rm nuc}$ and 19\% for $n=5n_{\rm
nuc}$ respectively). In any case, a hadron pressure always further
suppresses the rate for thermal nucleation at fixed $T$, $\mu$.
\bibitem{mardor} I.\ Mardor and B.\ Svetitsky, Phys.\ Rev.\ D {\bf 44},
878 (1991).
\bibitem{madsen} J.\ Madsen, Phys.\ Rev.\ D {\bf 50}, 3328 (1994).
\end{references}
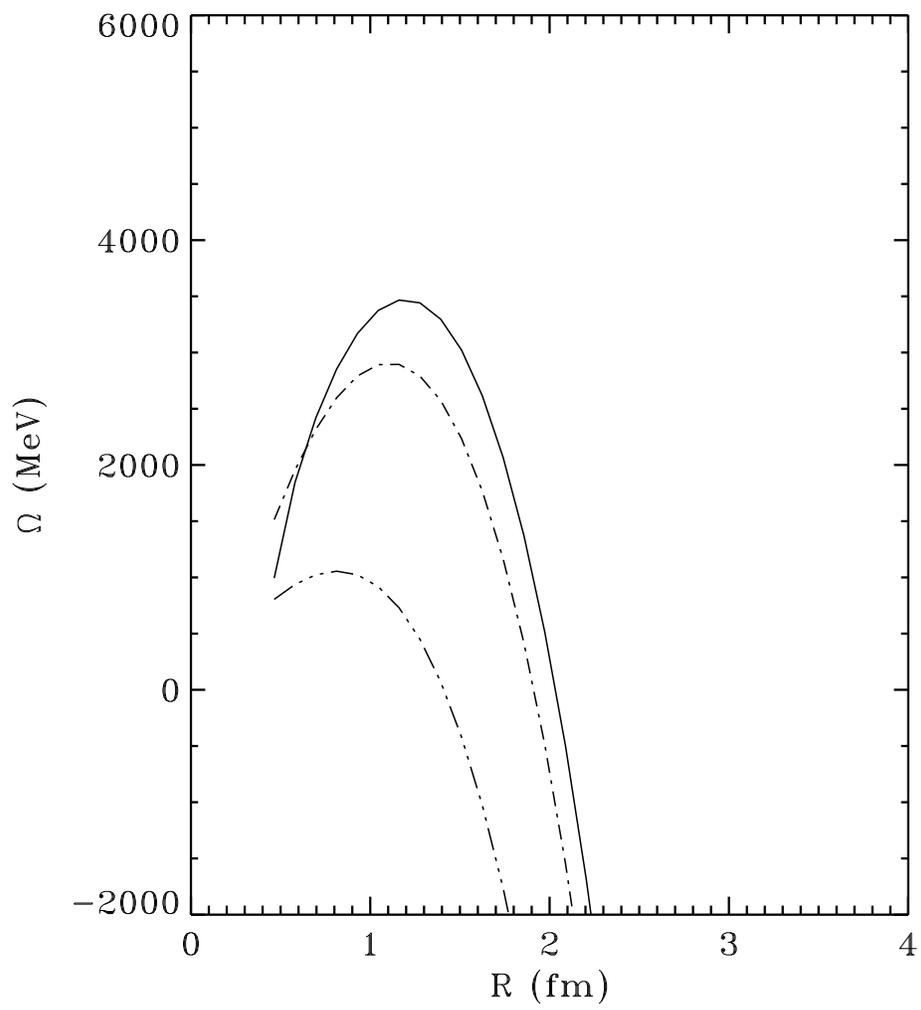
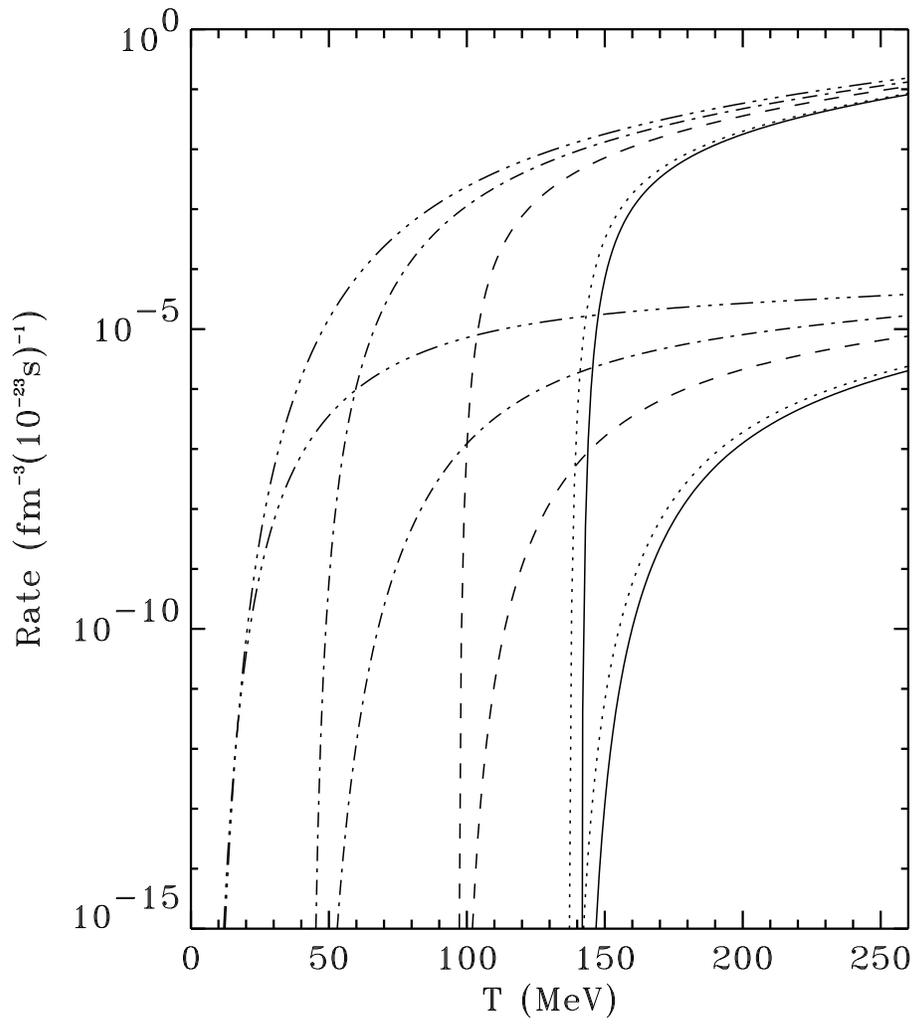
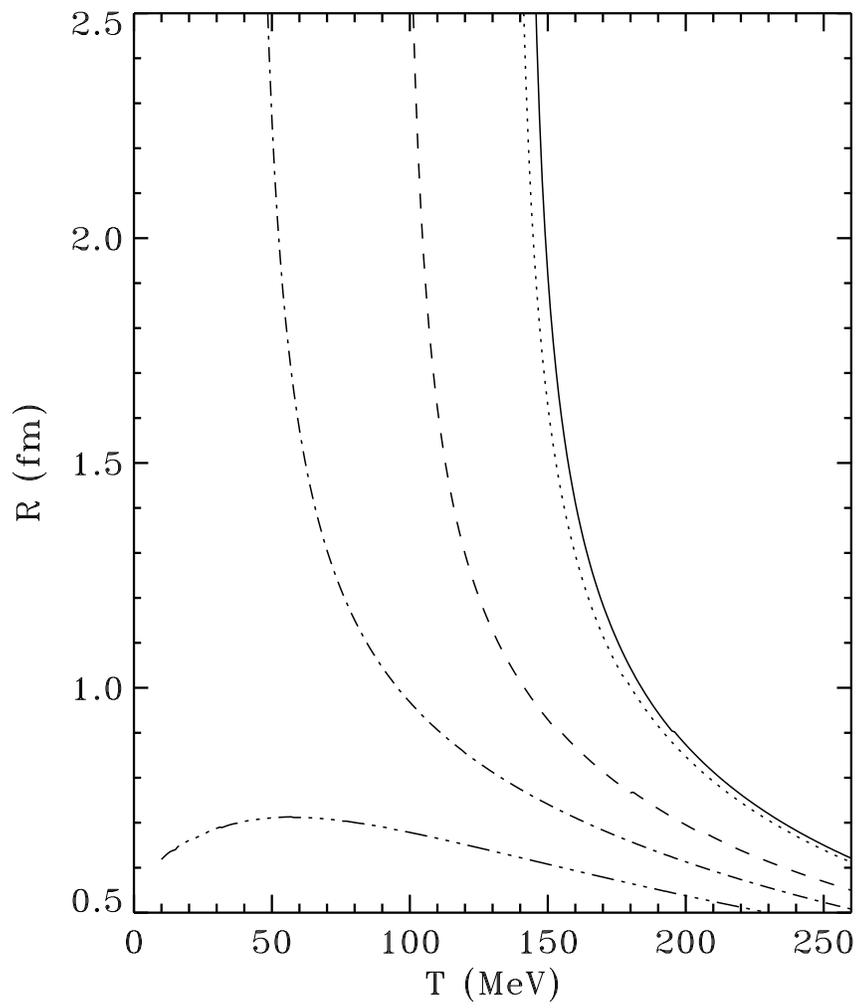
\end{document}